\documentclass[11pt, twocolumn]{article}

\usepackage[T1]{fontenc}
\usepackage[utf8]{inputenc}
\usepackage{authblk}
\usepackage{geometry}
\usepackage{graphicx}
\usepackage{amsmath}

\usepackage[font=small,labelfont=bf]{caption}
\title{\textbf{Phase-Tunable Thermal Logic: Computation with Heat}}

\author[1]{Federico Paolucci}
\author[1,2]{Giampiero Marchegiani}
\author[1]{Elia Strambini}
\author[1]{Francesco Giazotto}
\affil[1]{NEST, Instituto Nanoscienze-CNR and Scuola Normale Superiore, I-56127 Pisa, Italy}
\affil[2]{Dipartimento di Fisica dell'Universit\`a di Pisa, Largo Pontecorvo 3, I-56127 Pisa, Italy}

\begin{document}
\maketitle

\textbf{Boolean algebra, the branch of mathematics where variables can assume only true or false value, is the theoretical basis of classical computation. The analogy between Boolean operations and electronic switching circuits, highlighted by Shannon in 1938, paved the way to modern computation based on electronic devices. The grow of computational power of such devices, after an exciting exponential -Moore's trend, is nowadays blocked by heat dissipation due to computational tasks, very demanding after the chips miniaturization. Heat is often a detrimental form of energy which increases the systems entropy decreasing the efficiency of logic operations. Here, we propose a physical system able to perform thermal logic operations by reversing the old heat-disorder epitome into a novel heat-order paradigm. We lay the foundations of heat computation by encoding logic state variables in temperature and introducing the thermal counterparts of electronic logic gates. Exploiting quantum effects in thermally biased Josephson junctions (JJs), we propound a possible realization of a functionally complete "dissipationless" logic. Our architecture ensures high operation stability and robustness with switching frequencies reaching the GHz.}

\begin{figure}[ht]
	\includegraphics[width=\columnwidth]{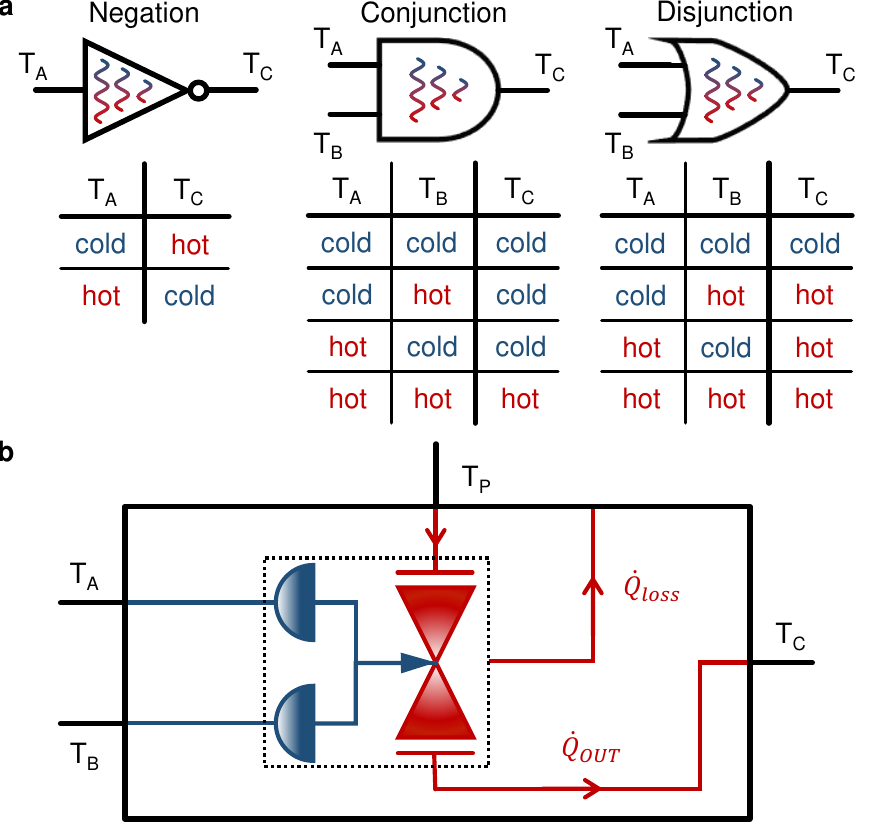}
	\caption{\textbf{Thermal Logic Gates.} \textbf{a}, Distinctive shapes, logic truth tables of NOT, AND and OR thermal logic gates ($cold$ is the logic state 0 while $hot$ is the logic state 1) are depicted. \textbf{b}, Schematic representation of a generic thermal logic gate. The device consists of two inputs (at temperature $T_A$ and $T_B$, respectively) controlling the heat flow between the power supply (at temperature $T_P$) and the output (at temperature $T_C$) through a valve (red hourglass) by means of an actuator (blue half-circumference). The output ($\dot{Q}_{OUT}$) and loss ($\dot{Q}_{loss}$) heat currents are shown.}
	\label{Fig1}
\end{figure}

The simplicity of Boolean algebra \cite{Boole1854} is one of the major motivations accounting for the development of classical computation during last century. The physical realization of mathematical operations defined in a binary environment [i.e. 0 (false) and 1 (true)] requires to employ systems where the variables can acquire only two stable values. Usually, this requirement is fulfilled through the use of electronic digital circuits implementing switching elements such as diodes and transistors \cite{Balabanian2001, Shannon1938, Bardeen1948}. Modern devices entrain control systems able to perform logic operations without passing through electronic interfaces or external calculus apparatus \cite{Groisman2003}. For this reason logic architectures taking advantage of different physical systems, such as: optics \cite{Kitayama1985}, fluidics \cite{Prakash2007, Travagliati2012}, pneumatics \cite{Jensen2007} and molecules \cite{Aviram1988}, have been developed. What makes a calculation scheme appealing for technological applications is the operation speed. The most popular approach is quantum computation \cite{Ladd2010}, where coherent quantum states exploit high-fidelity quantum bits (qubits) in order to implement computation algorithms \cite{Deutsch1985, Debnath2016, Prando2017}. Unfortunately, all logic architectures share one unescapable side effect: heat generation due to dissipation \cite{Keyes2005, Mannhart2010, Gibbs1902}. Even the most efficient computation architecture dissipates a minimum amount of heat estimated by Landauer fundamental limit \cite{Landauer1961}. Energy harvesting chases the storage and conversion of ambient energy into autonomous new functions. In this framework, recycling the already produced heat to perform logic operations would allow to recover part of lost power.

Here, we discuss the grounds of a thermal logic by constructing a functionally complete architecture \cite{Enderton2001} through the definition of thermal logic gates. The latter employ temperature as logic state variable which can acquire only two digital values: $cold$ (logic state 0) and $hot$ (logic state 1). A thermal logic device controls the energy flow between a reservoir and the output lead by means of a control mechanism (actuator) coupled to the input contacts. In such systems, the balance between the transmitted energy (which depends on the input temperatures configuration) and the power losses defines the output temperature (output logic state). The modulation of electron heat currents in solid-state nanostructures \cite{Giazotto2012, Mart2014} can be realized in the framework of coherent caloritronics \cite{Martinez2014, Meschke2006, Martinez2015, FornGiaz2017}. Heat currents are mastered by manipulating the superconducting quantum phases across thermally-biased Josephson junctions through the application of an external magnetic flux \cite{Martinez2013, Fornieri2015, Strambini2014, Fornieri2016}. The latter can be generated by taking advantage of another quantum effect: thermoelectricity in temperature-biased normal metal-ferromagnetic insulator-superconductor ($N-FI-S$) tunnel junctions \cite{Ozaeta2014, Kolenda2016}. In this Article, we show how the marriage of these two quantum effects develops a functionally complete thermal logic. The proposed structure guarantees full complementarity with low temperature electronic computating systems. In particular, the $N-FI-S$ junctions allow to re-convert the thermal signals into electrical ones. As a consequence, the design of hybrid thermal/electrical systems (where areas of the circuit perform electrical computation and other thermal calculation) is possible. This possibility could pave the way to new computation concepts and architectures.

\section*{Computation with Heat}
The three main operations of Boolean algebra \cite{Boole1854} are: negation ($NOT$), conjunction ($AND$) and disjunction ($OR$). All the other operations can be obtained by a composition of them, i.e. they define a functionally complete logic. Figure \ref{Fig1}-a carries the essence of thermal logic: the distinctive shapes and the related truth tables of negation, conjuction and disjunction thermal logic gates are shown. For instance, the output temperature of a $NOT$ thermal logic gate is 1 ($hot$) when the input is 0 ($cold$), and viceversa. The specific value of the two logic states $cold$ and $hot$ only depends on the technology chosen to realize the architecture.

The most generic concept of thermal logic gate is a device with one or more binary temperature inputs that set the binary output. Such a system consists of one or more thermal inputs that control the heat current flowing through a valve connecting a power supply to an output contact, as schematized in Figure \ref{Fig1}-b. The steady-state output temperature of such general device $T_C$, i.e the result of the logic operation, derives from the balance between the output and loss heat currents. Therefore, it can be calculated by solving the following equation:

{\footnotesize
\begin{equation}
	\label{Tbal}
		\dot{Q}_{OUT}\left(T_A, T_B, T_P, T_C\right)-\dot{Q}_{loss}\left(T_A, T_B, T_P, T_C\right)=0
\end{equation}
}
where $\dot{Q}_{OUT}$ is the heat current flowing through the thermal valve, $\dot{Q}_{loss}$ includes all the heat losses of the device, $T_A$ and $T_B$ are the input temperatures, and $T_P$ is the power supply temperature (always kept at $T_{hot}$). Note that the type of implemented logic gate ($OR$, $AND$ or $NOT$) is exclusively defined by the structure of the actuators.

In solid state structures, phonons are the capital carriers exclusively of heat, while electrons carry both heat and charge. Since phonons are difficult to control, up to now the advancements in phononics \cite{Li2012} are less shattering than in electronics. Therefore, the use of phonons for thermal logic operations \cite{Wang2007} is challenging. On the contrary, in coherent caloritronics \cite{Martinez2014, Meschke2006, Martinez2015, Martinez2013, Fornieri2015, Strambini2014, Fornieri2016} electron heat currents are precisely manipulated through the control of phases in superconducting mesoscopic circuits, such as superconducting quantum interferences devices ($SQUID$s) \cite{Giazotto2012, Mart2014} and superconducting quantum intereference proximity transistors ($SQUIPT$s) \cite{Strambini2014}. In such systems (i.e. metal thin nanostructures) the vanishing Kapitza resistance \cite{Giazotto2012} ensures that the phonons of every element are completely thermalized with the substrate at a constant bath temperature $T_{bath}$. At low temperatures (usually $T_{bath}<1$K) the electron-phonon coupling is weak \cite{Giazotto2006}. As a consequence, the electron and phonon ensembles are in thermal disequilibrium (i.e $T_e\neq T_{ph}$ with $T_e$ and $T_{ph}$ electron and phonon temperature, respectively) and the sum of the power losses due to the electron-phonon thermal gradient in each element of the device $\dot{Q}_{e-ph}$ is the principal cause of heat losses in the system ($\dot{Q}_{loss}=\sum_{i}\dot{Q}_{{e-ph},i}$). 

\section*{Phase-Tunable Thermal Logic}
An ideal logic architecture requires the input and output states to be identical variables (to allow for scalable networks) and fully decoupled (to avoid crosstalk).

Both requirements are fulfilled by the recently proposed first fully-thermal caloritronic device, the phase-tunable temperature amplifier ($PTA$) \cite{Paolucci2016}. It employs a thermal nano-valve (i.e. a temperature-biased $SQUIPT$ \cite{Strambini2014}) controlled by the magnetic flux $\Phi$ resulting from the closed-circuit current generated by a low temperature thermoelectric element (i.e. a $N-FI-S$ tunnel junction) \cite{Ozaeta2014, Kolenda2016, Giazotto2015} closed by a superconducting coil. This inductive coupling guarantees an almost infinite input-to-output impedance making the $PTA$ and ideal fully-thermal transistor when operated at unitary gain  \cite{Paolucci2016}.

\begin{figure}[t!]
	\includegraphics[width=\columnwidth]{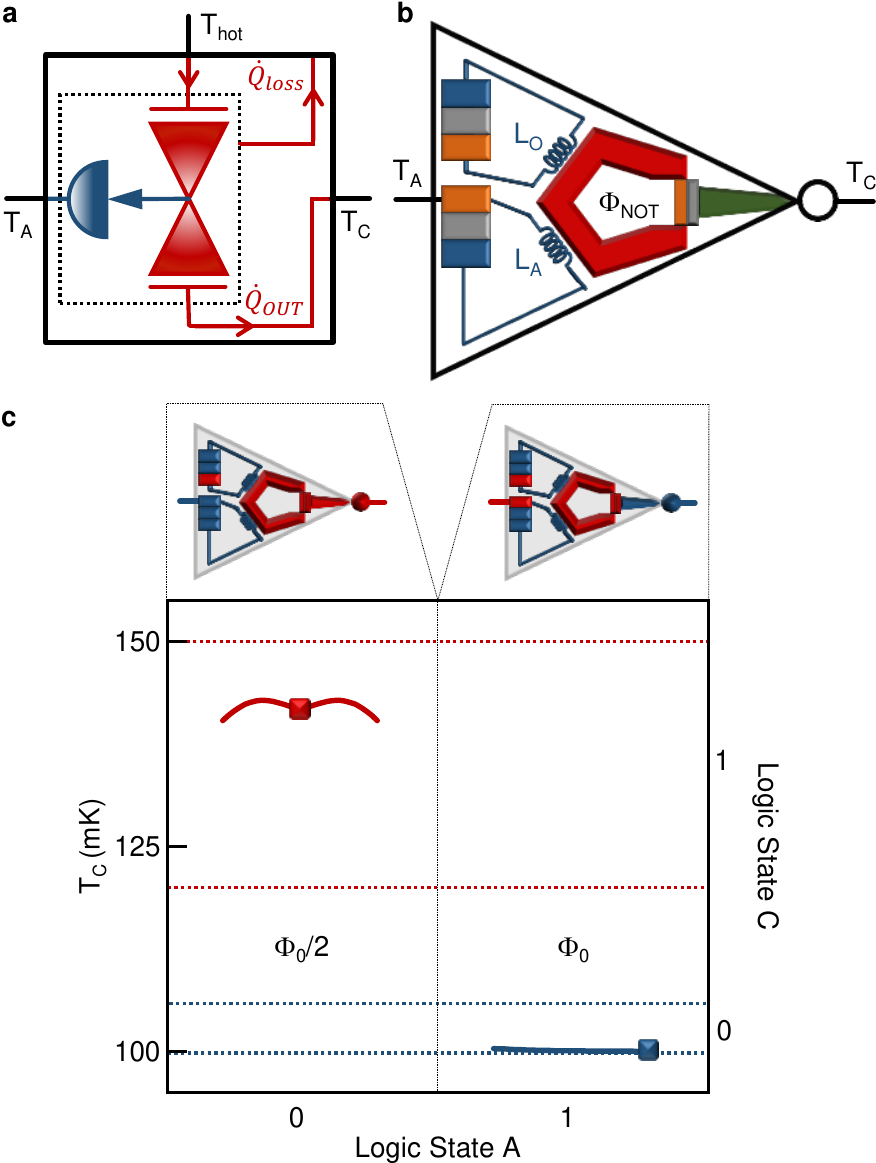}
	\caption{\textbf{NOT Logic Gate.} \textbf{a}, Thermal schematic: an actuator (blue half-circumference) at input temperature $T_A$ controls through an normally open valve (red hourglass) the heat flow from the power supply (at temperature $T_{hot}$) to the output (at temperature $T_C$). The output ($\dot{Q}_{OUT}$) and loss ($\dot{Q}_{loss}$) thermal currents are shown. \textbf{b}, Schematic of the coherent caloritronic realization: thermoelectric elements are constituted of a metal (orange), a ferromagnetic insulator (gray) and a superconductor (blue). The blue spirals depict the superconducting coils $L_A$ and $L_O$. The $SQUIPT$ is composed of a superconductor ring (red) interrupted by a metal wire (orange) tunnel-coupled to a metal probe (green) through a thin insulator (gray). \textbf{c}, Output temperature $T_C$ and output logic state $C$ versus the logic input configuration for $T_{cold}=T_{bath}=100~$mK and $T_{hot}=150~$mK. The output temperature $T_C$ for optimum (squares) and fluctuating (lines) input is shown.}
	\label{Fig2}
\end{figure}

In the following we briefly introduce the temperature-biased $SQUIPT$ and the $N-FI-S$ junction, and we show how the synergy between these two building blocks can be used to implement a functionally complete thermal logic architecture.

The $SQUIPT$ is composed of a superconducting ring interrupted by a normal metal wire. The latter acquires a superconducting character through the superconducting proximity effect. A normal metal tunnel probe acting as output lead is coupled to the wire \cite{Giazotto2010}. A magnetic flux $\Phi$ threading the ring modulates the density of states ($DOS$) of the proximized wire \cite{Petrashov1995, leSueur2008} and, as a consequence, the thermal conductance between the wire and the tunnel probe is periodic with the magnetic flux with period $\Phi=\Phi_0$ \cite{Strambini2014}. In particular, the heat current is minimum (smaller than the power losses due to electron-phonon coupling) for $\Phi=0$ (when  the full superconducting minigap is developed in the wire $DOS$) and maximum for $\Phi=\Phi_0/2$ (when the wire shows a normal metal $DOS$). A detailed description of the thermal nano-valve ($SQUIPT$) can be found in the Methods section. 

On the other hand, a thermoeletric effect can be generated by breaking the electron-hole symmetry in the $DOS$ of a conductor. In a superconductor, this condition can be accomplished by Zeeman spin-splitting the $DOS$ through an exchange field and spin-filtering the quasiparticles \cite{Ozaeta2014}. Both the mechanisms can be provided by a single ferromagnetic insulator layer of a $N-FI-S$ junction. A temperature gradient between the normal metal and the superconductor produces a thermoelectric signal: an open circuit thermovoltage $V_T$ in the Seebeck regime or a closed circuit thermocurrent $I_T$ in the Peltier regime \cite{Ozaeta2014, Kolenda2016, Giazotto2015}. A detailed description of the thermoelectric element can be found in the Methods. 

The controlling system (actuator) is realized by shorting the thermoelectric element with a superconducting coil that inductively controls the thermal valve, as depicted in Figure \ref{Fig1}-b. This allows to design different "dissipationless" thermal logic gates, because the energy used for the logic operation comes from otherwise lost power, and the dissipation in the actuation electric system is zero (the current flows in a superconducting coil).

\begin{figure*}[ht]
  \centering
	\includegraphics[width=0.75\textwidth]{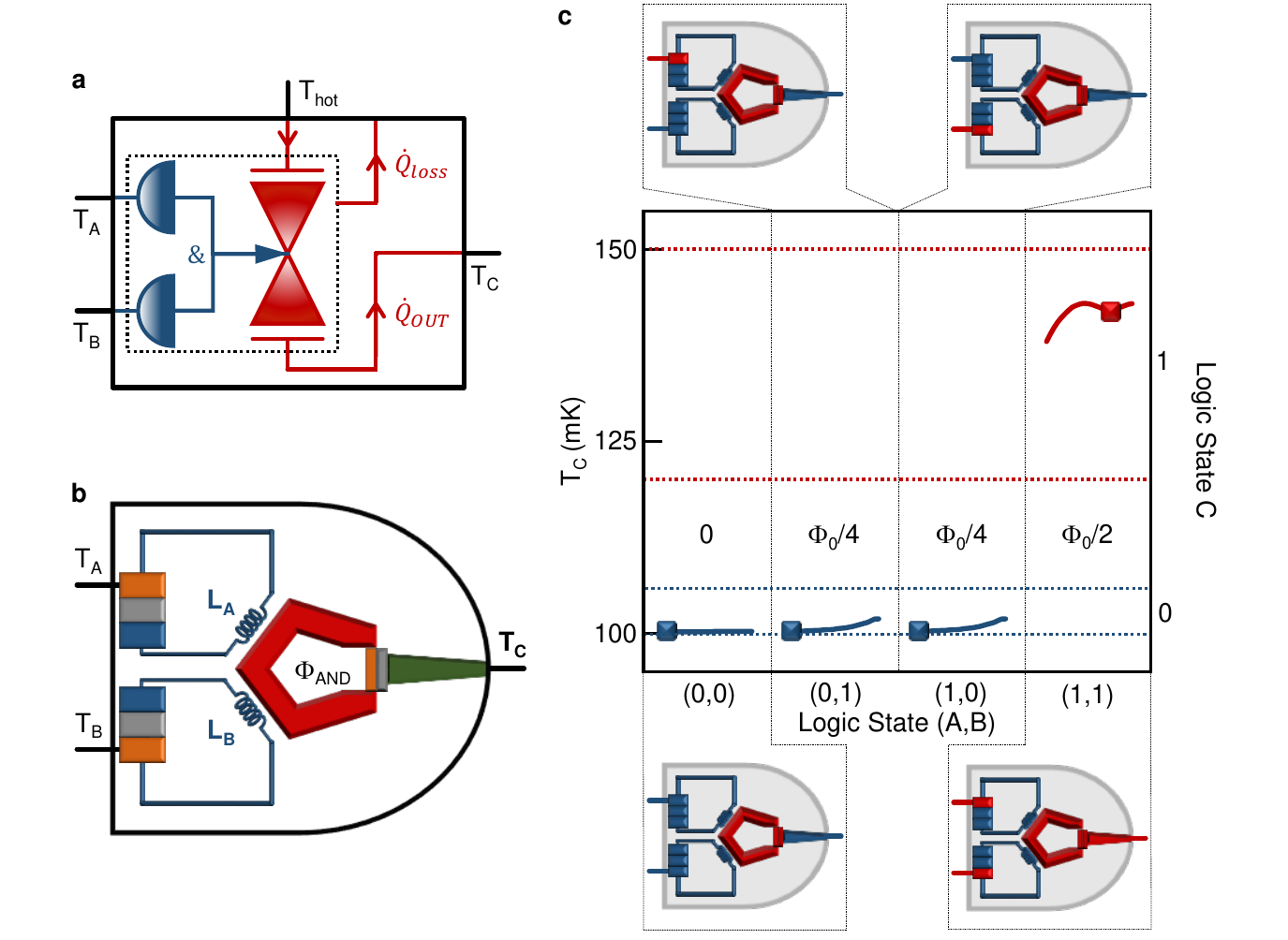}
	\caption{\textbf{AND Logic Gate.} \textbf{a}, Thermal schematic: two actuators (blue half-circumferences) at input temperature $T_A$ and $T_B$ control through a valve (red hourglass) the heat flow between the power supply (at temperature $T_{hot}$) and the output (at temperature $T_C$). The output ($\dot{Q}_{OUT}$) and loss ($\dot{Q}_{loss}$) thermal currents are shown. \textbf{b}, Schematic of the coherent caloritronic realization: thermoelectric elements are constituted of a metal (orange), a ferromagnetic insulator (gray) and a superconductor (blue). The blue spirals depict the superconducting coils $L_A$ and $L_B$. The $SQUIPT$ is composed of a superconductor ring (red) interrupted by a metal wire (orange) tunnel-coupled to a metal probe (green) through a thin insulator (gray). \textbf{c}, Output temperature $T_C$ and output logic state $C$ versus the logic input configuration for $T_{cold}=T_{bath}=100~$mK and $T_{hot}=150~$mK. The output temperature $T_C$ for optimum (squares) and fluctuating (lines) input is shown.}
	\label{Fig3}
\end{figure*}

\subsection*{Negation Logic Gate - NOT}
The thermal inverter logic gate $NOT$ can be in outline conceived as a normally open valve, as depicted in Figure \ref{Fig2}-a. In this way, when $T_A=T_{cold}$ (input logic state 0) a thermal current flows through the valve and the output temperature is $T_C=T_{hot}$ (output logic state 1). The energization of the actuator interrupts the flow of heat current from the power supply to the output lead and, as a consequence, the output temperature $T_C$ lowers to $T_C=T_{cold}=T_{bath}$ (output logic state 0). This can be realized by controlling a temperature-biased $SQUIPT$ (thermal valve) through two $N-FI-S$ junctions shorted by two coils: one connected to the input $T_A$ while the other always kept at $T_{hot}$ (see Figure \ref{Fig2}-b). The total magnetic flux threading the superconducting ring $\Phi_{NOT}(T_A)$ is the sum of the contributions due to the opening $\Phi_{O}$ and the input $\Phi_{A}$ coil, and it takes the form:

{\footnotesize
\begin{equation}\label{eq:PhiNot}
\begin{split}
&\Phi_{NOT}(T_A)=\Phi_{O}(T_{hot})+\Phi_{A}(T_A)=\\
&=M_O I_{T,O}(T_{hot})+M_A I_{T,A}(T_A),\\
\end{split}
\end{equation}
}
where $M_{i}$ (with $i=O, A$) is the mutual inductance between the superconducting ring and the opening $L_O$ or input coil $L_A$, and $I_{T,i}$ is the thermocurrent generated by the opening or input thermoelectric element. The mutual inductance $M_i$ is chosen in order to have maximum conduction through the valve when $T_i=T_{hot}$, i.e $\Phi_{i}(T_{hot})=\Phi_{0}/2$ (with $i=O,A$). In summary, the behavior of the $NOT$ thermal logic gate is expressed by the following system:

{\footnotesize
\begin{equation}
\label{TabNot}
\begin{aligned}
\Phi_{NOT}(T_A)=
  \begin{cases}
    \frac{\Phi_{0}}{2}+0       & \text{if }T_{A}=T_{cold} \implies T_{C}=T_{hot}\\
    \frac{\Phi_{0}}{2}+\frac{\Phi_{0}}{2}  & \text{if }T_{A}=T_{hot}\text{ } \implies T_{C}=T_{cold}.\\
  \end{cases}
\end{aligned}
\end{equation}
}
The opening coil $L_O$ provides a constant contribution $\Phi_{0}/2$ to the flux. As consequence, when $T_{A}=T_{cold}$ the input coil does not give any contribution to the total flux and the $SQUIPT$ conducts ($T_{C}=T_{hot}$), while for $T_{A}=T_{hot}$ the total flux is $\Phi_{NOT}=\Phi_0$ and the heat current through the valve is almost completely suppressed ($T_{C}=T_{cold}$).

\begin{figure*}[ht]
  \centering
	\includegraphics[width=0.75\textwidth]{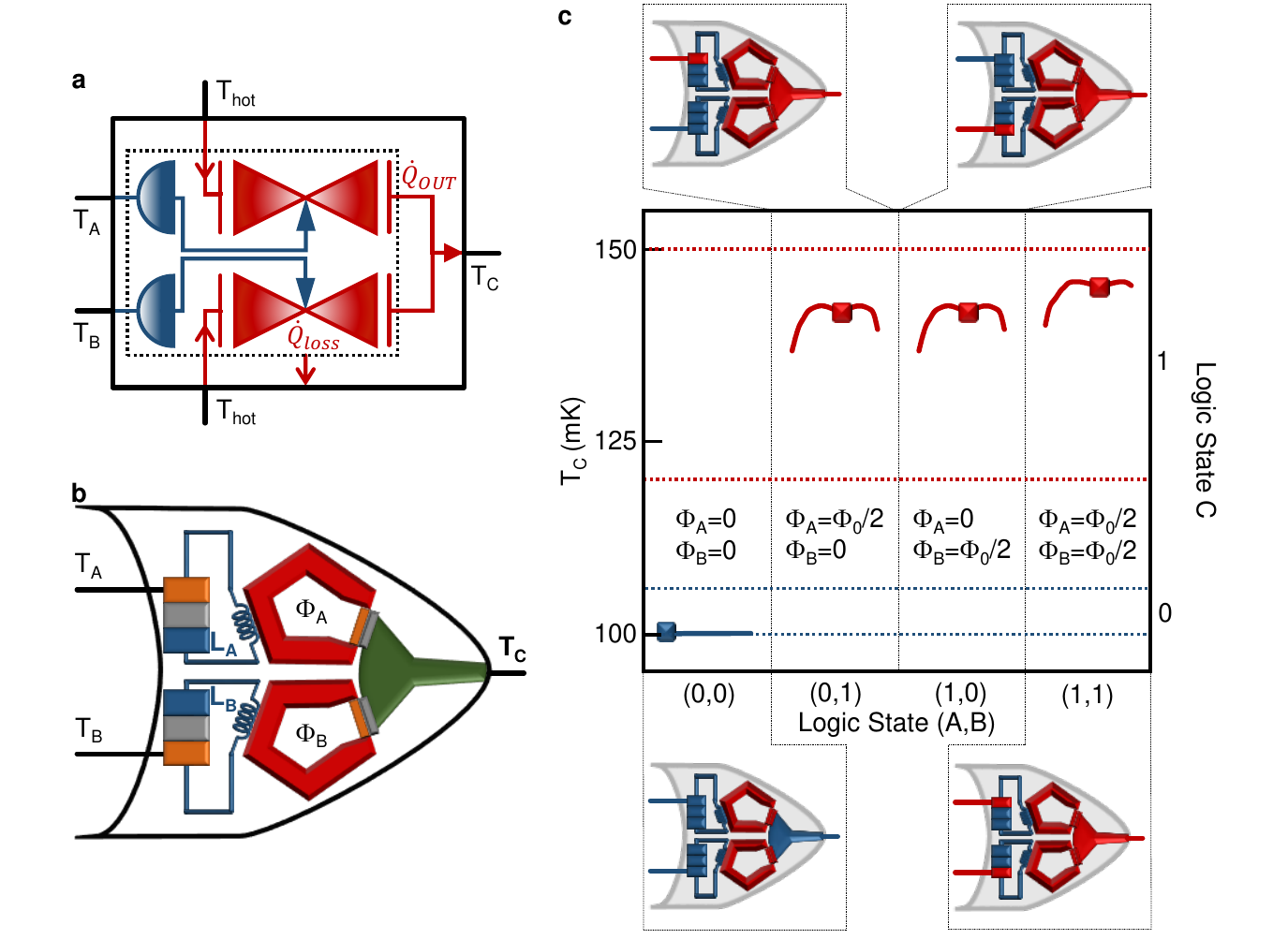}
	\caption{\textbf{OR Logic Gate.} \textbf{a}, Thermal schematic: two actuators (blue half-circumferences) at input temperature $T_A$ and $T_B$ control through two valves (red hourglasses) the heat flow from the power supply (at temperature $T_{hot}$) to the output (at temperature $T_C$). The output ($\dot{Q}_{OUT}$) and loss ($\dot{Q}_{loss}$) thermal currents are shown. \textbf{b}, Schematic of the coherent caloritronic realization: thermoelectric elements are constituted of a metal (orange), a ferromagnetic insulator (gray) and a superconductor (blue). The blue spirals depict the superconducting coils $L_A$ and $L_B$. The $SQUIPT$s are composed of a superconductor ring (red) interrupted by a metal wire (orange) tunnel-coupled to a metal probe (green) through a thin insulator (gray). \textbf{c}, Output temperature $T_C$ and output logic state $C$ versus the logic input configuration for $T_{cold}=T_{bath}=100~$mK and $T_{hot}=150~$mK. The output temperature $T_C$ for optimum (squares) and fluctuating (lines) input is shown.}
	\label{Fig4}
\end{figure*}

In our setting, we define as logic state 0 the temperatures ranging from 100 to $105~$mK (with optimum value $T_{cold}=100~$mK) and logic state 1 the temperatures in the range $120-150~$mK (with optimum value $T_{hot}=150~$mK). In order to demonstrate the feasibility of our architecture we employ a Cu-EuS-Al tunnel-junction \cite{Giazotto2015} as thermoelectric element and an Al-based $SQUIPT$ with a Cu output electrode as thermal nano-valve \cite{Strambini2014, Giazotto2010}. Further details on materials and geometry can be found elsewhere \cite{Paolucci2016}. 

The output characteristic of the thermal inverter is shown in Figure \ref{Fig2}-c. For an ideal logic input 0 ($T_{A}=100~$mK) the output temperature is $T_C=142~$mK, while for logic input 1 ($T_{A}=150~$mK) the output becomes $T_C=100~$mK. Fluctuations on the order of $10\%$ around the optimum input signal value produce variations of $T_C$ that do not compromise the correct operation of the system as shown in the same figure. 

\subsection*{Conjunction Logic Gate - AND}

The thermal conjuction logic gate $AND$ can be represented as a valve which is always closed, except when both actuators are energized ($T_C=T_{hot}$ only if $T_A=T_B=T_{hot}$), as depicted in Figure \ref{Fig3}-a. In the coherent caloritronic realization, the heat flow across a temperature-biased $SQUIPT$ is modulated by means of the total magnetic flux $\Phi_{AND}(T_A, T_B)$ generated by the coils $L_A$ and $L_B$ shorting two thermoelectric elements (input leads) at temperature $T_A$ and $T_B$. The total magnetic flux threading the superconducting ring is the sum of the contribution due to the two inputs ($\Phi_A$ and $\Phi_B$) and is given by:

{\footnotesize
\begin{equation}\label{eq:PhiAnd}
\begin{aligned}
&\Phi_{AND}(T_A, T_B)=\Phi_{A}(T_{A})+\Phi_{B}(T_B)=\\
&=M_A I_{T,A}(T_{A})+M_B I_{T,B}(T_B),\\
\end{aligned}
\end{equation}
}
where $M_{i}$ (with $i=A,B$) is the mutual inductance between the superconducting ring and the input coils $L_i$, and $I_{T,i}$ is the thermocurrent generated by the input thermoelectric elements. Both mutual inductances $M_i$ are chosen in order to have $\Phi_{i}(T_{hot})=\Phi_{0}/4$ (with $i=A,B$). As a consequence, the behavior of the $AND$ thermal logic gate can be summarized by the following system:

{\footnotesize
\begin{equation}
\label{TabAnd}
\begin{aligned}
\Phi_{AND}(T_A, T_B)=\\
  \begin{cases}
    0+0       & \text{if }T_{A}=T_{cold}\text{, } T_{B}=T_{cold} \implies T_{C}=T_{cold}\\
    0+\frac{\Phi_{0}}{4}       & \text{if }T_{A}=T_{cold}\text{, } T_{B}=T_{hot}\text{ } \implies T_{C}=T_{cold}\\
    \frac{\Phi_{0}}{4}+0       & \text{if }T_{A}=T_{hot}\text{, }\text{ } T_{B}=T_{cold} \implies T_{C}=T_{cold}\\
    \frac{\Phi_{0}}{4}+\frac{\Phi_{0}}{4}       & \text{if }T_{A}=T_{hot}\text{, }\text{ } T_{B}=T_{hot}\text{ } \implies T_{C}=T_{hot}.
  \end{cases}
\end{aligned}
\end{equation}
}
When $T_{A}=T_B=T_{cold}$ [input logic state $(A,B)=(0,0)$] the input coils do not generate any magnetic flux ($\Phi_{AND}=0$), thereby the thermal conductance of the $SQUIPT$ is almost zero ($T_{C}=T_{cold}$) (output logic state 0). If a single actuator is active, i.e. $T_A=T_{hot}$ or $T_B=T_{hot}$ with the other input at $T_{cold}$, the total magnetic flux driving the $SQUIPT$ is $\Phi_{AND}=\Phi_0/4$. Therefore, the nano-valve is still almost completely closed \cite{Strambini2014, Paolucci2016} and $T_{C}=T_{cold}$ (output logic state 0). For $T_{A}=T_B=T_{hot}$ the total magnetic flux is $\Phi_{AND}=\Phi_0/2$ and the $SQUIPT$ fully conducts. As a consequence, the output logic state is 1 ($T_{C}=T_{hot}$).

For the numerical demonstration of the behavior of the thermal conjunction we employ the same materials and geometry used for the $NOT$ gate \cite{Paolucci2016}. Figure \ref{Fig3}-c illustrates the transfer characteristic of the $AND$ logic gate. In the case of partially conducting $SQUIPT$ [i.e. for $(A,B)=(1,0)$ or $(A,B)=(0,1)$], a small heat current flows from the power supply to the output electrode and the $T_C$ is slightly larger than for $(A,B)=(0,0)$. In all cases, the output temperature resides within the range of logic state 0 ($100-105~$mK), because the electron-phonon coupling is large enough to partially compensate the effect of the heat current reaching the output (see Methods).

\subsection*{Disjunction Logic Gate - OR}

The working principle of the thermal disjunction logic gate $OR$ is resumed in Figure \ref{Fig4}-a, where the parallel connection of two normally closed valves ($T_C=T_{cold}$ when $T_A=T_B=T_{cold}$) allows the heat flow from the power supply at temperature $T_P=T_{hot}$ to a common output electrode. The heating of at least one actuator ($T_A=T_{hot}$ and/or $T_B=T_{hot}$) opens a conduction channel to the output ($T_C=T_{hot}$). A possible practical realization is constituted of two $SQUIPT$s sharing the same output electrode, as schematized in Figure \ref{Fig4}-b. Each thermal nano-valve is controlled by the magnetic flux ($\Phi_A$ or $\Phi_B$) generated by a thermoelectric element connected to the input electrode (at temperature $T_A$ or $T_B$). The magnetic flux $\Phi_{OR}$ effectively controlling the $OR$ logic gate can be defined as:

{\footnotesize
\begin{equation}\label{eq:PhiOr}
\begin{aligned}
&\Phi_{OR}(T_A, T_B)=\Phi_{A}(T_{A})\lor \Phi_{B}(T_B)=\\
&=M_A I_{T,A}(T_{A})\lor M_B I_{T,B}(T_B),\\
\end{aligned}
\end{equation}
}
where $M_{i}$ (with $i=A,B$) is the mutual inductance between the superconducting ring and the input coil $L_i$, and $I_{T,i}$ is the thermocurrent generated by the input thermoelectric elements. The mutual inductance $M_i$ is chosen in order to have $\Phi_{i}(T_{hot})=\Phi_{0}/2$ (with $i=A,B$). In summary, the $OR$ thermal logic gate works as follows:

{\footnotesize
\begin{equation}
\label{TabOr}
\begin{aligned}
\Phi_{OR}(T_A, T_B)=\\
  \begin{cases}
    0\lor0       & \text{if }T_{A}=T_{cold}\text{, } T_{B}=T_{cold} \implies T_{C}=T_{cold}\\
    0\lor\frac{\Phi_{0}}{2}       & \text{if }T_{A}=T_{cold}\text{, } T_{B}=T_{hot}\text{ } \implies T_{C}=T_{hot}\\
    \frac{\Phi_{0}}{2}\lor0       & \text{if }T_{A}=T_{hot}\text{, }\text{ } T_{B}=T_{cold} \implies T_{C}=T_{hot}\\
    \frac{\Phi_{0}}{2}\lor\frac{\Phi_{0}}{2}       & \text{if }T_{A}=T_{hot}\text{, }\text{ } T_{B}=T_{hot}\text{ } \implies T_{C}=T_{hot}.
  \end{cases}
\end{aligned}
\end{equation}
}
For $T_{A}=T_B=T_{cold}$ [input logic state $(A,B)=(0,0)$], both input coils do not generate any magnetic flux ($\Phi_{OR}=0$), therefore both $SQUIPT$s are shut ($T_{C}=T_{cold}$) and the output logic state is 0. When at least one actuator is active, i.e. $T_A=T_{hot}$ or $T_B=T_{hot}$, there is heat flowing to the output electrode and $T_{C}=T_{hot}$ (output logic state 1).

\begin{figure}[t!]
	\centering
	\includegraphics[width=\columnwidth]{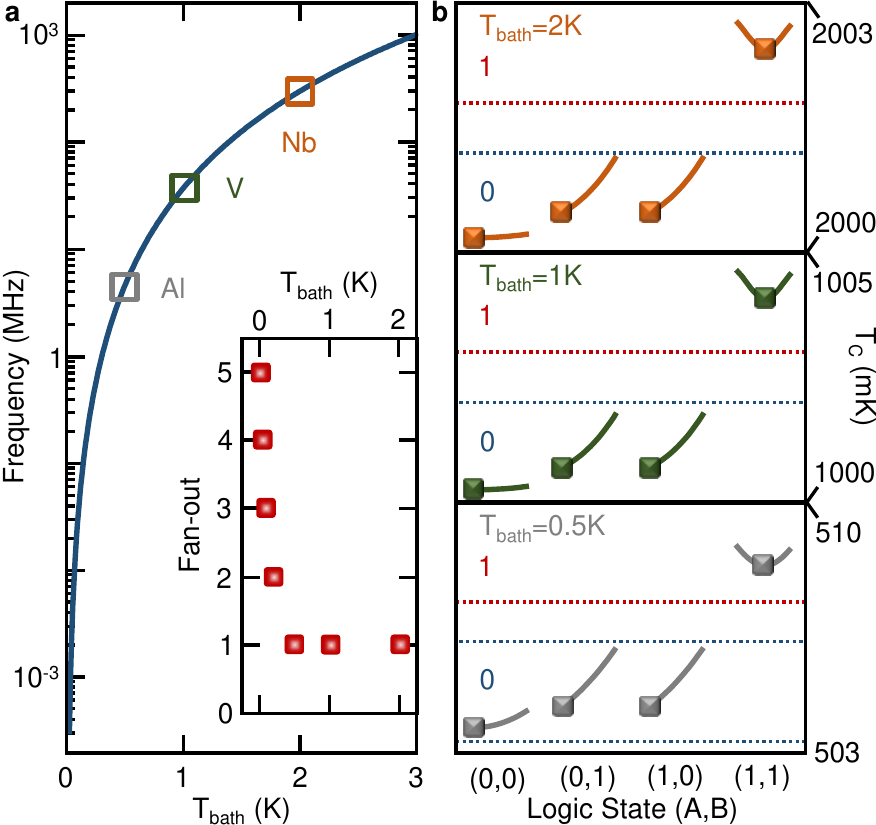}
	\caption{\textbf{Temperature dependence of the logic architecture.} \textbf{a}, Maximum operation frequency as a function of the bath temperature $T_{bath}$. Inset: fan-out as a function of the phonon temperature $T_{bath}$. \textbf{b}, Output Temperature $T_C$ versus the input configuration $(A,B)$ of an $AND$ logic gate for $T_{bath}=0.5, 1, 2~K$. The output signal for optimum (squares) and fluctuating (lines) input is shown.}
	\label{Fig5}
\end{figure}

The behavior of the thermal disjunction is numerically determined by employing the same materials and geometry used for negation and conjuction gates \cite{Paolucci2016}. The transfer characteristic of the $OR$ logic gate is resumed in Figure \ref{Fig4}-c. When only one actuator is energized [i.e. for $(A,B)=(1,0)$ or $(A,B)=(0,1)$], only the heat current flowing throug one valve reaches the output electrode and the optimal output temperature is $T_C=142~$mK (for further details see the Methods). In the case of $(A,B)=(1,1)$, the thermal current flowing through both $SQUIPT$s arrive at the output; as a consequence, the output temperature is higher than in the previous cases ($T_C=145~$mK).

\section*{Operation temperature, speed, fan-out and compatibility}

Temperature is the fundamental working parameter for phase-tunable thermal logic, because the grounding physical mechanisms of coherent caloritronics, such as electron-phonon coupling and superconducting pairing potential, strongly depend on both phonon and electron temperatures. For example, the working speed of such logic gates is limited by the thermalisation of the electrons with the lattice phonons, because typically the characteristic time constant of the inductive coupling is much shorter. Therefore, at first approximation, the logic working speed depends on the threshold temperature $T_{cold}=T_{bath}$ \cite{Giazotto2006}. In the limit of clean metals (employed as output electrode in our architecture) the electron-phonon relaxation time is given by \cite{Giazotto2006}:

{\footnotesize
\begin{equation}\label{eq:time}
\tau_{e-ph}=\frac{k_B^2v_F}{0.34\Sigma T_{bath}^3}
\end{equation}
}
where $k_B$ is the Boltzmann constant, $v_F$ is the Fermi velocity, and $\Sigma$ is the electron-phonon coupling constant of the metal output lead. Therefore, the maximum operation frequency ($f=1/\tau_{e-ph}$) is limited by the phonon temperature, as depicted in Figure \ref{Fig5}-a. For $T_{cold}=100~$mK (which was used up to now), the operation frequency $f$ is limited to about $100~$KHz. By employing superconductors with higher critical temperature, such as vanadium ($T_{V}=5.38$K) and niobium ($T_{Nb}=9.3$K), the operation temperature can be increased, and $f$ reachesvalues on the order of a few GHz.

Although we demontrated a thermal logic architecture working at temperatures higher than $1~$K (see Figure \ref{Fig5}-b), for logic state 1 the output temperature is drastically suppressed compared to $T_{hot}$, and the separation between $T_{cold}$ and $T_{hot}$ results to be very small (a few mK). We introduce the fan-out, namely the number of series logic gates properly working without the necessity of signal amplification (see the inset of Figure \ref{Fig5}-a). It is calculated by using the output heat of a device (in logic state 1) as input signal for the next logic gate until $T_C$ for state 1 resides in the correct temperature range. At high temperature every logic gate requires an amplification of the output signal, therefore the temperature amplifier \cite{Paolucci2016} becomes integral part of the device. On the contrary, at low temperatures the fan-out rises. For example, in the case of $T_{cold}=100~$mK it is possible to connect, in principle, 3 devices before amplification. 

Here, we have demonstrated the proof of principle of a new phase-tunable thermal logic by employing the simplest and most common geometry in hybrid nanostructures. However, the performances of our architecture (speed and fan-out) can be drastically improved by using only superconducting materials, because the thermal losses due electron-phonon coupling drastically decrease \cite{Giazotto2006}. For instance, fully superconducting thermal memories working up to about $10~$K at frequencies up to tens of GHz have been proposed \cite{Guarcello2017}. In order to speed up our system and increase the fan-out, thermoeletric elements based on $S-FI-S'$ tunnel junctions (with $\Delta_S > \Delta_{S'}$) \cite{Giazo2015} and fully superconducting temperature-biased $SQUIPT$s \cite{Strambini2014} could be employed.

Finally, we would like to highlight that phase-tunable thermal logic could be used in synergy with other computation approaches. It can utilize the unavoidable heat generated by dissipation in other logic architectures in order to increase the total calculating capacity and to decrease the energy consuption. For instance, the employed materials and geometry are fully compatible with standard low temperature semiconductor-based technologies and quantum computation architectures. Therefore, phase-tunable thermal logic could represent a fertile field for the growth of new and more efficient combined computation systems.

\section*{Methods}
\subsection*{Thermal valve - SQUIPT}
Electronic thermal currents flowing from a power supply to an output electrode through a tunnel barrier are given by \cite{Giazotto2006}:
{\footnotesize
\begin{equation}
\begin{split}
&\dot{Q}_{OUT}(T_P,T_{C})=\\
&=\frac{2}{e^2R_{T}}\int_{0}^{\infty}N_{P}(E) N_{C}(E)\left\lbrack f_0(E,T_P)-f_0(E, T_{C}) \right\rbrack E dE
\end{split}
\label{eq:Jsquipt}
\end{equation}
}
where $T_{P/C}$ is the temperature of the power supply or the output lead, $e$ is the electron charge, $R_T$ is the normal state tunnel resistance, $N_{P/C}$ is the reduced DOS for the power supply or the output lead and $f_0(E, T)=\left\lbrack1+\exp (E/k_BT)\right\rbrack^{-1}$ is the Fermi distribution of the quasiparticles. 

The thermal valve is a device which controls the flow of a heat current by opening and closing a passageway. The thermal current is modulated by tuning the $DOS$ of (at least) one of the two electrodes \cite{Strambini2014}. For simplicity, in the following we assume an output electrode made of a normal metal with $N_{C}(E)=1$ and we tune the DOS of the power supply electrode $N_{P}(E)$. This can be realized by placing a normal metal wire in good electric contact with a superconducting ring ($S$). The superconducting properties acquired by the wire through the proximity effect \cite{Holm1932} can be modulated by a magnetic flux $\Phi$ threading the superconducting loop \cite{Giazotto2010, Petrashov1995, leSueur2008, Meschke2011}. The DOS of the wire $N_{P}=\left\vert\Re\left\lbrack g^R \right \rbrack\right\vert$ is the real part of the retarded Green's function $g^R$ \cite{Rammer1986} obtained by solving the one-dimensional Usadel equation \cite{Usadel1970}. In the limit of short junction (i.e. when $E_{Th}=\hbar D/l^2\gg \Delta_{0_{S}}$, where $E_{Th}$ is the Thouless energy, $\hbar$ is the reduced Planck constant, $D$ is the wire diffusion coefficient, $l$ is the length of the wire and $\Delta_{0_{S}}$ is the zero-temperature superconducting energy gap of the ring) the proximity effect is maximized, and the DOS can be explicitly written \cite{Strambini2014, Giazotto2010}:
{\footnotesize
\begin{equation}
\begin{split}
N_{P}(E,\Phi)=\left| \Re\left\lbrack \frac{E-iE_{Th}\gamma g_s}{\sqrt{(E-iE_{Th}\gamma g_s)^2+\left\lbrack E_{Th}\gamma f_s \cos \left( \frac{\pi \Phi}{\Phi_0} \right) \right\rbrack^2}} \right\rbrack \right|.
\end{split}
\label{eq:DOSwire}
\end{equation}
}
Above, $\gamma=R_{P}/R_{int}$ is the transmissivity of the $S-P$ contact (where $R_{P}$ is the resistance of the metal wire and $R_{int}$ the resistance of the $S-P$ interface), $g_S(E)=\frac{E+i\Gamma_{S}}{\sqrt{(E+i\Gamma_{S})^2-\Delta_{S}^2}}$ and $f_S(E)=\frac{\Delta_{S}}{\sqrt{(E+i\Gamma_{S})^2-\Delta_{S}^2}}$ are the coefficients of the phase-independent and phase-dependent parts of proximized DOS (where $\Gamma_{S}$ is the Dynes broadening parameter \cite{Dynes1984} and $\Delta_{S}$ is the BCS energy gap \cite{Tinkham}), and $\Phi_0\simeq2.0678\times10^{-15}~$Wb is the magnetic flux quantum. The periodic behavior of $N_{C}(E,\Phi)$ in the magnetic flux (with periodicity $\Phi=\Phi_0$) results in a heat current $\dot{Q}_{OUT}(T_P,T_{C})$ with the same periodic dependence on $\Phi$.

For a given supply temperature $T_P$, the steady-state temperature of the output electrode $T_C$ is obtained by solving the following energy balance equation:
{\footnotesize
\begin{equation}
\begin{split}
-\dot{Q}_{OUT}(T_P, T_{C},\Phi)+\dot{Q}_{e-ph, C}(T_{C}, T_{bath})=0.
\end{split}
\label{eq:BalEq}
\end{equation}
}
The electron-phonon coupling takes the form $\dot{Q}_{loss, C}=\dot{Q}_{e-ph, C}(T_{C}, T_{bath})=\Sigma~ V \left(T_{C}^n-T_{bath}^n\right)$, where $\Sigma$ is the electron-phonon coupling constant, $V$ is the volume of the output eletrode and the exponent $n$ depends on the disorder of the system \cite{Giazotto2006}. For metals, in the clean limit $n=5$, while in the dirty limit $n=4,6$ \cite{Strambini2014}. From Equation \ref{eq:BalEq} descends that the temperature on the right side of the tunnel junction $T_{C}$ inherits the same dependence on $\Phi$ of $N_{C}$ and $\dot{Q}_{OUT}$ (i.e. $T_{C}$ shows a minimum for $\Phi=0$ and maximum around $\Phi=\Phi_0/2$).

\subsection*{Actuation system}
Since the nano-valve ($SQUIPT$) is controlled by a magnetic flux, it is necessary a temperature-to-flux conversion mechanism in the actuation system. This is realized by a thermoelectric element shorted by a superconducting coil. 

Electron-hole asymmetry in the quasiparticle $DOS$ is the key ingredient for thermoelectricity \cite{Mermin}. In superconductors it can be accomplished by spin-splitting the $DOS$ through an exchange field $h_{ex}$ and by selecting a specific spin species through the coupling of the superconductor to a spin-polarized element \cite{Ozaeta2014}. Both requirements are satisfied by a normal metal-ferromagnetic insulator-superconductor ($N-FI-S$) junction, where the ferromagnetic element produces both the exchange field $h_{ex}$ and the polarization $P=(G_{\uparrow}-G_{\downarrow})/(G_{\uparrow}+G_{\downarrow})$ [where $G_{\uparrow}$ and $G_{\downarrow}$ are the spin up and spin down conductances] \cite{Moodera2007, Stramb2017}. For a superconductor thinner than the coherence length $\xi_0$, the spin-splitted $DOS$ can be assumed to be spatially homogeneous \cite{Tokuyasu1988} and written \cite{Giaz2008}:
{\footnotesize
\begin{equation}
\begin{split}
N_{\uparrow,\downarrow}(E)=\frac{1}{2} \left| \Re\left\lbrack \frac{E+i\Gamma \pm h_{ex}}{\sqrt{(E+i\Gamma \pm h_{ex})^2-\Delta^2}} \right\rbrack \right|,
\end{split}
\label{eq:DOSsplit}
\end{equation}
}
where $E$ is the energy, $\Gamma$ is the Dynes broadening parameter, and $\Delta(T, h_{ex})$ is the superconducting order parameter, which is calculated self-consistently from the $BCS$ equation \cite{Giaz2008}:
{\footnotesize
\begin{equation}
\begin{split}
\ln \left(\frac{\Delta_0}{\Delta} \right)=\int_0^{\hbar\omega_D}\frac{f_+(E,T)+f_-(E,T)}{\sqrt{E^2+\Delta^2}}dE.
\end{split}
\label{eq:GapEq}
\end{equation}
}
Above, $\Delta_0$ is the zero-temperature superconducting gap, $\omega_D$ is the Debye frequency of the superconductor and {\footnotesize$f_{\pm}(E,T)= \left \{ 1+\exp\left\lbrack\left( \sqrt{E^2+\Delta^2}\mp h_{ex}\right)/k_BT \right\rbrack \right \}^{-1}$} is the Fermi distribution of the spin-polarized electrons. 

The thermocurrent originated by keeping $N$ at a temperature $T_A$ and the other two elements ($FI$ and $S$) at the bath temperature $T_{bath}$ takes the form:
{\footnotesize
\begin{equation}
\begin{split}
&I_T(T_A, T_{bath})=\\
&=\frac{1}{eR_T}\int_{-\infty }^{\infty }\left\lbrack N_+(E)+PN_-(E)\right\rbrack \left\lbrack f_N(E,T_{A})-f_S(E,T_{bath}) \right\rbrack dE,
\end{split}
\label{eq:It}
\end{equation}
}
where $R_T$ is the tunnel resistance in the normal state, $N_{\pm}(E)=N_{\uparrow}(E)\pm N_{\downarrow}(E) $ and $f_{N,S}(E,T_{A,bath})= \left \lbrack 1+\exp\left(E/k_BT_{A,bath} \right)\right \rbrack^{-1}$ is the Fermi distribution of the metal or the superconductor. The  resulting magnetic flux which threads the superconducting ring is:
{\footnotesize
\begin{equation}
\begin{split}
&\Phi(T_A, T_{bath})=MI_T(T_A, T_{bath})=k\sqrt{L_AL_S}I_T(T_A, T_{bath}),
\end{split}
\label{eq:phi}
\end{equation}
}
where $M$ is the mutual inductance, $k\leq1$ is the coupling coefficient, $L_A$ is the inductance of the coil shorting the $N-FI-S$ junction and $L_S$ is the geometric inductance of the superconducting ring.

\section*{Acknowledgements}
The authors thank A. Braggio for the useful discussions.
The authors acknowledge the European Research Council under the European Unions Seventh Framework Programme (FP7/2007-2013)/ERC Grant No. 615187 - COMANCHE for partial financial support. The work of F.P. is funded by Tuscany Region under the FARFAS 2014 project SCIADRO. The work of E.S. is funded by a Marie Curie Individual Fellowship (MSCA-IFEF-ST No. 660532-SuperMag).

\section*{Author contributions statement}
F.P. and F.G. conceived the architecture, F.P., G.M. and E.S. developed the model, F.P wrote the manuscript.  All authors reviewed the manuscript. 

\section*{Additional information}
The authors declare no competing financial interests.
\end{document}